\documentstyle[psfig,preprint,tighten,aps]{revtex}

\def\bea{\begin{eqnarray}}
\def\eea{\end{eqnarray}}

\def\nn{\nonumber}
\def\Tr{{\,\mbox{Tr}}}

\begin{document}
%\draft
\preprint{TAN-FNT-00-01}

\title{Non-leptonic hyperon weak decays in the Skyrme model revisited}

\author{Daniel G\'omez Dumm$^a$, Andrea J. Garc{\'{\i}}a$^a$ and Norberto N. Scoccola$^{a,b}$ }

\address{
$^a$ Physics Department, Comisi\'on Nacional de Energ{\'{\i}}a At\'omica, \\
      Av.Libertador 8250, (1429)~Buenos Aires, Argentina.\\
$^b$ Universidad Favaloro, Sol{\'\i}s 453, (1078) Buenos Aires, Argentina.}

\maketitle

\begin{abstract}
Non-leptonic hyperon weak decays are investigated in the $SU(3)$
Skyrme model. We use a collective coordinate scheme,
following the approach in which the symmetry breaking terms in
the strong effective action are diagonalized exactly. To
describe the weak interactions we use an octet dominated weak
effective lagrangian that leads to a good description of the known
$2\pi$ and $3\pi$ kaon decays. We show that the observed $S$-wave
decays are reasonably well reproduced in the model. On the other
hand, our calculated $P$-wave amplitudes do not agree
with the empirical ones even though both pole and
contact contributions to these amplitudes are properly taken into
account. Finally, an estimate of the non-octet contributions to
the decay amplitudes is presented.
\end{abstract}

\vspace{1cm}

%\pacs{PACS number(s): 12.39.Dc, 13.30.Eg, 12.39.Fe}

\section{Introduction}

Although during the last few decades much progress has been done in the theoretical study of hadron structure, the
non-leptonic weak decays of hyperons still remain far from being well understood. This class of decays involve not
only weak interactions but also low momentum strong processes, which have made their first principle calculation
unfeasible so far. In this situation, different hadron models have been used to get the corresponding theoretical
predictions. The available experimental information allows to determine both $S$- and $P$-wave decay amplitudes
separately for various processes. In the case of the $S$-wave decays, the predictions given by quark models with
QCD enhancement factors turn out to be quite successful (see Ref.\cite{DGH86} and references therein). These
models, however, have serious difficulties in reproducing at the same time the empirical results in the case of the
$P$-wave amplitudes. Indeed, this seems to be a problem (so called ``$S/P$ wave puzzle'') which is common to other
approaches as e.g.\ heavy baryon chiral perturbation theory\cite{Jen92},
QCD sum rules\cite{HHK99}, etc.

A possible solution to the $S/P$ wave puzzle has been suggested some
years ago within the context of chiral topological soliton
models\cite{DGH86,DGL85}. It was shown that in these models the
$P$-wave amplitudes receive, in addition to the standard pole diagrams,
extra contributions from contact terms. Then, it was speculated
that these extra terms could provide a clue to this issue.
Unfortunately, at the time this suggestion was made such models were
hampered by several serious problems, such as very poor predictions for the
hyperon spectrum\cite{Che85}, far too small results for the
$S$-wave non-leptonic decay amplitudes\cite{DGL85,PT85}, etc.
Therefore, it was hard to draw definite conclusions about the real
relevance of the contact terms. With the introduction of more
refined methods to treat the chiral symmetry breaking terms in the
effective action the situation has significantly improved. Indeed,
a scheme in which one introduces $SU(3)$ collective coordinates
and the hamiltonian ---including a symmetry breaking piece--- is
diagonalized exactly leads to very good results for the hyperon
spectra, as well as reasonable predictions for different baryon
properties (for a review see Ref.\cite{Wei96}). Moreover, it has
been recently shown\cite{Sco98} that within such scheme, and using a
simple Cabbibo current-current form for the weak interaction, one
can obtain the correct $S$-wave absolute values.
Thus, we are now in position to verify whether chiral soliton models
can provide a unified and consistent description of the hyperon
non-leptonic decays. In this work we will describe the weak
interactions by means of an effective weak chiral lagrangian ---which
is more general than the Cabbibo current-current
coupling often used in previous Skyrme model
calculations \cite{PT85,Sco98,TF86,KSO90}---, where
the low energy coupling constants will be fixed to
reproduce the known $2\pi$ and $3\pi$ $K$-meson decays. We will
concentrate mostly in the dominant octet-like piece of this
$\Delta S=1$ lagrangian, considering terms up to order $p^4$.
In addition, possible non-octet contributions will be considered
for the particular case of the ${\cal A}(\Sigma^+_+)$ amplitude
which, as well known, vanishes in the pure octet approximation.
The non-leptonic hyperon decay amplitudes will be obtained by evaluating
the corresponding matrix elements using the topological soliton model
wave functions.

The article is organized as follows: in Sec.\ II we give a brief overview
of the $SU(3)$ soliton model and introduce the octet-like weak effective
chiral lagrangian to be used in the following two sections. In
Sec.\ III and IV we describe the calculation of the $S$-wave
and $P$-wave amplitudes respectively, and present the corresponding
results. The impact of the non-octet-like components of the weak
effective lagrangian is discussed in Sec.\ V and in Sec.\ VI we
state our conclusions. Finally, in the Appendix we give some
details concerning the evaluation of the matrix
elements of the collective operators which appear in the calculation
of the decay amplitudes.

\section{The model}

In chiral topological soliton models baryons are described as topological excitations
of a chiral effective action which depends only on meson fields.  We use the form
\begin{equation}
\Gamma = \Gamma_{SK} + \Gamma_{WZ} + \Gamma_{SB} \ ,
\label{action}
\end{equation}
where $\Gamma_{SK}$ and $\Gamma_{WZ}$ stand for the Skyrme
and Wess-Zumino actions respectively, and $\Gamma_{SB}$ is
an $SU(3)$ symmetry breaking piece. The Skyrme action has the
usual form
\begin{equation}
\Gamma_{SK} =
\int d^4 x \Big\{ {f^2_\pi \over 4}
\Tr\left[ \partial_\mu U (\partial^\mu U)^\dagger \right]
+
 {1\over{32 \epsilon^2}}
 \Tr\left[ [U^\dagger \partial_\mu U , U^\dagger \partial_\nu U]^2\right] \Big\}
 \, ,
\end{equation}
where the chiral field $U$ is a non--linear realization of the
pseudoscalar octet, $f_\pi = 93$ MeV is the pion decay constant
and $\epsilon$ is the dimensionless Skyrme
parameter. The Wess-Zumino action reads
\begin{eqnarray}
\Gamma_{WZ} &=& - {i N_c \over{240 \pi^2}} \int d^5x
\,\epsilon^{\mu\nu\rho\sigma\tau} \Tr\,[ L_\mu L_\nu L_\rho L_\sigma
L_\tau]\, ,
\end{eqnarray}
where $L_\mu = U^\dagger \partial_\mu U$ and $N_c=3$ is the number
of colors. Finally the symmetry breaking piece $\Gamma_{SB}$ is given
by
\begin{eqnarray}
\Gamma_{SB} & = &\int d^4x \left\{
 { f_\pi^2 m_\pi^2 + 2 f_K^2 m_K^2 \over{12} }
 \Tr \left[ U + U^\dagger - 2 \right]
%\right. \nonumber \\
%& & \qquad \left.
+ \sqrt{3}  { f_\pi^2 m_\pi^2 - f_K^2 m_K^2 \over{6} } \Tr \left[
\lambda_8 \left( U + U^\dagger \right) \right] \right. \nonumber
\\ & & \qquad \left. + { f_K^2 - f_\pi^2\over{12} } \Tr \left[ (1-
\sqrt{3} \lambda_8) \left( U (\partial_\mu U)^\dagger \partial^\mu
U + U^\dagger \partial_\mu U (\partial^\mu U)^\dagger \right)
\right] \right\} \, ,
\label{sb}
\end{eqnarray}
where $f_K$ is the kaon decay constant and $m_\pi$ and $m_K$ are
the pion and kaon masses, respectively. In our numerical
calculations below we will set these parameters to their empirical values
and take $\epsilon = 4.1$ which is suitable for a good
description of many baryon properties in this model.

In the soliton picture we are using the strong interaction
properties of the low--lying $\frac{1}{2}^+$ and $\frac{3}{2}^+$
baryons are computed following the standard $SU(3)$ collective
coordinate approach to the Skyrme model. We introduce for the
chiral field the ansatz
\begin{equation}
U_0({\bf r}, t) = A(t) \ \left(
\begin{array}{cc}
c\ \openone + i \mbox{\boldmath $\tau$} \cdot
{\hat{\mbox{\boldmath $r$}}} \ s \ & 0 \\ 0 & 1
\end{array}
\right) \ A^\dagger(t) \ ,
\label{ansatz}
\end{equation}
where we have used the abbreviations $c= \cos F(r)$ and $s=\sin
F(r)$, $F(r)$ being the chiral angle that parameterizes the
soliton. The collective rotation matrix $A(t)$ is $SU(3)$ valued.
Substituting the configuration Eq.\ (\ref{ansatz}) into $\Gamma$
yields (upon canonical quantization of $A$) the collective
Hamiltonian. Its eigenfunctions are identified as
the baryon wavefunctions $\Psi_B$. Due to
the symmetry breaking piece
$\Gamma_{SB}$, the Hamiltonian is obviously not $SU(3)$ symmetric.
However, as shown by Yabu and Ando \cite{YA88}, it can be
diagonalized exactly. The diagonalization essentially amounts to
admixtures of states from higher $SU(3)$ representations
into the octet ($J=\frac{1}{2}$) and decouplet
($J=\frac{3}{2}$) states. This procedure has proven to be quite
successful in describing the hyperon spectrum and static
properties \cite{Wei96}.

In order to describe the non-leptonic hyperon decays we have to
introduce an effective weak $\Delta S=1$ lagrangian. The latter
is constrained by weak interactions to transform either as
$\underline{8}$ or $\underline{27}$ under the chiral group $SU(3)_L$.
Here, we will take into account only the dominant octet-like couplings,
which lead to pure $\Delta I=1/2$ transitions. The remaining
$\underline{27}$ piece includes both $\Delta I=1/2$ and $\Delta I=3/2$
operators and turns out to be suppressed in view of the yet not completely
understood ``$\Delta I=1/2$ rule". Further considerations about these
non-octet-like terms will be given in Sec.\ V. We consider the
effective lagrangian given by~\cite{DGH84}
\begin{equation}
{\cal L}^{(8)}_w = g \Tr\,[ \lambda_6 \partial_\mu U
\partial^\mu U^\dagger ] +
      g' \Tr\,[ \lambda_6 \partial_\mu U \partial^\mu U^\dagger
        \partial_\nu U \partial^\nu U^\dagger] +
     g'' \Tr\,[ \lambda_6 \partial_\mu U \partial_\nu U^\dagger
     \partial^\mu U \partial^\nu U^\dagger]\ .
\label{weaklag}
\end{equation}
It should be noticed that this is not the most general octet-like $\Delta
S=1$ interaction one can write down up to ${\cal O}(p^4)$ in the
momentum power expansion. The latter, containing many
other terms, has been presented in Ref.\cite{KMW90}. For the decays we
are interested in (no
external fields), it turns out that the most general lagrangian
includes 15 independent terms of order $p^4$ leading to pure
$\Delta I=1/2$ transitions. However, to this order, it has been
shown\cite{KMW91} that the couplings considered in (\ref{weaklag})
are sufficient to fit the known data on
$K\to\pi\pi$ and $K\to\pi\pi\pi$. In the absence of further
information from the meson sector, we will just stick to this
simple form. In order
to give an idea of the uncertainties in our calculations, we will
consider two sets of values for the constants $g,g'$ and $g''$
which provide fits to the kaon data of similar quality.
Set A corresponds to the parameters used in Refs.~\cite{DGL85,DGH84},
\begin{equation}
g = 3.60 \times 10^{-8} \ m_\pi^2 \quad ; \quad
g'/g = 1.50 \times 10^{-1} \ {\rm fm}^2 \quad ; \quad
g''/g = -6.74 \times 10^{-2} \ {\rm fm}^2\;,
\label{donval}
\end{equation}
while Set B corresponds to the values obtained in Ref.\cite{KMW91},
\begin{equation}
g = 2.98 \times 10^{-8} \ m_\pi^2 \quad ; \quad
g'/g = 1.69 \times 10^{-1} \ {\rm fm}^2 \quad ; \quad
g''/g =  1.87 \times 10^{-2} \ {\rm fm}^2\;.
\label{kamval}
\end{equation}

To calculate the hyperon decays in the context of the Skyrme model
with $SU(3)$ collective coordinates, we include the soft meson
fluctuations on top of the soliton background. This is achieved
using
\begin{equation}
U = U_M \ U_0({\bf r}, t) \ U_M \ ,
\label{fullu}
\end{equation}
where $U_M = 1 + i \vec \tau \cdot \vec \pi /(2\, f_\pi) +...$.
Inserting this expression into ${\cal L}^{(8)}_w$ and taking the
appropriate matrix elements one can obtain the desired
$S$-wave and $P$-wave decay amplitudes. This is worked out
in the following two sections.

\section{$S$-wave amplitudes}

As mentioned in the Introduction, it has been recently
shown\cite{Sco98} that if the Cabbibo current-current form is
used to describe the weak interactions,
the present soliton model leads to a
reasonably good description of the $S$-wave hyperon decay amplitudes.
In this section we study these amplitudes using the effective
weak chiral lagrangian given by Eqs.~(\ref{weaklag}--\ref{kamval}).

As usual, we assume that isospin symmetry is preserved. In such
limit, the following relations between the non-leptonic decay
amplitudes can be derived:
\begin{mathletters}
\begin{eqnarray}
\Sigma^-_- &=& \Sigma_+^+ -\sqrt2 \ \Sigma^+_0
\label{samp}\\
\Lambda_-^0 &=& -\sqrt2\ \Lambda^0_0 \\ \Xi_-^- &=& - \sqrt2\
\Xi^0_0 \ ,
\end{eqnarray}
\end{mathletters}
\hspace{-.15cm}where the lower indices indicate the charge of
the outgoing pion.
In this way, only four of the seven measurable amplitudes need to
be considered as independent. For simplicity, we choose these
amplitudes to be $\Lambda^0_0$, $\Sigma^+_0$, $\Xi^0_0$ and
$\Sigma_+^+$.

For a process $B \rightarrow B'\pi$, we can define the amplitudes
${\cal A}$ and ${\cal B}$, corresponding to $S$- and $P$-wave decays
respectively, according to
\begin{equation}
i\, \langle B' \pi | {\cal L}^{(8)}_w | B \rangle = \bar
u_{B'}\,({\cal A}+{\cal B}\,\gamma_5)\, u_B \ .
\end{equation}
In the soft-pion limit, the octet nature of ${\cal L}^{(8)}_w$, together
with current algebra relations, lead to an additional constraint
for the $S$-wave $\Sigma$ decay amplitudes, namely ${\cal A}(\Sigma^-_-)=
-\sqrt{2}\,{\cal A}(\Sigma^+_0)$. Thus from (\ref{samp}) one obtains
${\cal A}(\Sigma^+_+)=0$. The remaining ${\cal A}$ amplitudes
can be calculated by replacing Eqs.\ (\ref{fullu})
and (\ref{ansatz}) in ${\cal L}^{(8)}_w$ and taking the corresponding
matrix elements. We find
\begin{equation}
{\cal A}(B \rightarrow B' \pi^0) = \alpha  \ \langle B'| R_{78} | B\rangle\,,
\label{amp}
\end{equation}
where
\begin{equation}
\alpha = \frac{4 \pi\, i}{\sqrt{3} f_\pi}  \int dr \ r^2 \left[ g
\left( F'^2 + 2 \ \frac{\sin^2 F}{r^2} \right)   - g'  \left( F'^2
+ 2 \ \frac{\sin^2 F}{r^2} \right)^2  - g'' \left( F'^4 - 4 \
\frac{F'^2 \sin^2 F}{r^2} \right) \right]\,.
\end{equation}
In Eq.\ (\ref{amp}), $R_{78}$ stands for an $SU(3)$ rotation matrix,
$R_{ab} = 1/2 \ \Tr\left[ \lambda_a A \lambda_b A^\dagger \right]$.
As explained in the Appendix, its matrix elements between the collective
wavefunctions describing the different baryon states can be calculated
as linear combinations of $SU(3)$ Clebsch-Gordan coefficients.

Our results for the ${\cal A}$ amplitudes are summarized in
Table \ref{tasw}, where we also quote
the values corresponding to the quark model (QM)~\cite{DGH86}. Following
the usual convention, the overall phase has been fixed to obtain
${\cal A}(\Lambda^0_-)$ real and positive. It can be seen from the
table that the predictions obtained with Set~A are about 15 \%
higher than those arising from Set~B. In both cases, our results
are somewhat below the experimental values. However, since
the deviation is in the same direction for all processes (notice that
this is not the case for the QM values), the agreement is significantly
improved if one considers the ratios between the different amplitudes.
In general, it could be said that our results and those corresponding to
the QM are of similar quality.

It is also interesting to compare the present results with those
of previous soliton calculations. As already mentioned, in some of
them\cite{PT85,Sco98,TF86,KSO90} the Cabbibo current-current has been
used. Within such scheme the best agreement with empirical data
has been obtained in Ref.\cite{Sco98} where, as done here, the baryon
wavefunctions arise from an exact diagonalization
of the SU(3) collective hamiltonian.
Generally speaking the results reported in Table~\ref{tasw} are somewhat
smaller (in absolute value) than those of Ref.\cite{Sco98}. However,
it should be stressed that the present calculation is free from the
uncertainties related to the question of whether (and to which extent)
QCD enhancement factors have to be included in soliton calculations.
Here, such factors are already accounted for in the value of the low
energy constants that appear in the weak lagrangian. On the other hand,
if our results are compared with previous calculations
based on effective weak chiral lagrangians\cite{DGL85}, we see that
the use of empirical input parameters in the strong effective
action, together with the exact diagonalization of the SU(3) collective
hamiltonian, lead to a significant improvement in the predictions.

A final comment concerns Ref.\cite{Gol87} and its Addendum
\cite{Gol87b}. In  Ref.\cite{Gol87}, the author studies weak
decays of hyperons in a chiral topological soliton model, starting
with a $\Delta S=1$ lagrangian including six ${\cal O}(p^4)$
terms. As shown in the Addendum, the Cayley-Hamilton theorem can
be used to reduce these six terms to only four independent ones.
Moreover, following the same steps as in our calculation, one
arrives to further relations between the corresponding hyperon
decay amplitudes. At the end one is effectively left with only two
terms, which ---as done in the present work--- can be chosen to be
those proportional to $g'$ and $g''$ in Eq.\ (\ref{weaklag}).
Following Ref.\cite{Gol87}, one might try to see whether it is
possible to find a set of coefficients for those terms capable to
reproduce the empirical values for both kaon and hyperon decays.
For this purpose, and in order to relate the coefficients with the
hyperon amplitudes, the author of Ref.\cite{Gol87} makes use of
the relation (\ref{amp}) and takes the $SU(3)$ symmetric limit. In
this way all decay amplitudes can be expressed in terms of one of
them and the $f/d$ ratio. The use of empirical values for the
latter leads to an incompatible system of equations, so that the
author argues that this precludes a successful application of the
chiral lagrangian model. We believe that the results found in the
present work provide some ground to disagree with such strong
conclusion. In fact, the relations in Ref.\cite{Gol87} are
expected to be modified by $SU(3)$ breaking effects and by
next-to-leading order corrections in $N_c$, which in general
introduce modifications to Eq.\ (\ref{amp}). From the results
displayed in Table~\ref{tasw} it is seen that if the chiral
soliton approach is used together with an effective weak chiral
lagrangian consistent with $K$ meson decays, one can obtain a
reasonably good description of the $S$-wave hyperon decays,
already at leading order in $N_c$.

\section{$P$-wave amplitudes}

We turn now to the evaluation of the $P$-wave amplitudes. In the
non-relativistic limit, they can be calculated using
\begin{equation}
{\cal B}\,\bar u_{B'}\,\gamma_5\, u_B
\simeq - {{\cal B}\over{2 \overline{M}}}
\ \chi^\dagger(\lambda')\ \vec \sigma \cdot \vec q \ \chi(\lambda)\ ,
\end{equation}
where $\overline{M}$ is the average of the $B$ and $B'$ empirical masses. As
stated in the Introduction, in the present model ${\cal B}$ receives
contributions of two different kinds\cite{DGL85}. One of them arises
from contact terms in ${\cal L}^{(8)}_w$ (see Fig.\ 1a), whereas the
other one is given by the pole diagrams shown in Figs.\ 1b-1d. Our
evaluation of the contact contribution in the case of $\pi^0$
emission leads to
\begin{equation}
{\cal B}_{contact}(B \rightarrow B' \pi^0) = -\frac{2 \overline{M}}{3 f_\pi}
\ \langle B'| \int d^3r\ {\hat {\cal P}}_c | B \rangle \,,
\end{equation}
where\footnote{Eq.\ (\ref{pcont})
shows some differences with respect to Eq.\ (24) of the erratum
of Ref.\cite{DGL85} which, we believe, contains errors and/or misprints.}
\begin{eqnarray}
\label{pcont}
{\hat{\cal  P}}_c & = & \quad g \left\{ (1+c) (F'+ 2 \frac{s}{r}) R_{63} -
                   \frac{2}{\sqrt{3}} (1-c) (F'- 2 \frac{s}{r}) R_3^{(+)} \right\}
\nonumber \\
           &   & +\; g' \left( F'^2 + 2 \frac{s^2}{r^2}\right) \left\{ - (1+c) (F'+ 2 \frac{s}{r}) R_{63} +
                       \frac{2}{\sqrt{3}}
                     \left[ F'(3 - c) + 2 \frac{s}{r} (3 c - 1) \right] R_3^{(+)} \right\}
\nonumber \\
           &   & +\; g'' \left\{ (1+c) F' \left( - F'^2 + 2 \frac{s}{r} F'+ 2 \frac{s^2}{r^2} \right) R_{63} \right.
\nonumber \\
           &   &
\qquad \qquad   \left.
                + \frac{2}{\sqrt{3}}
      \left[ F'^3 (3-c) + 2 \frac{s}{r} F'(1-c)  (F'- \frac{s}{r}) + 4 \frac{s^3}{r^3} c \right] R_3^{(+)}
                \right.
\nonumber \\
           &   &  \left.
\qquad \qquad
               - \frac{8}{\sqrt{3}} \frac{s}{r}
                 \left[ c F'^2 + F' \frac{s}{r} + c \frac{s^2}{r^2} \right] R_3^{(-)}
                 \right\} \;,
\end{eqnarray}
with
\begin{equation}
R_3^{(\pm)} =  R_{68} R_{33} \pm R_{63} R_{38}\;.
\label{rpm}
\end{equation}
Similar expressions can be found in the case of charged outgoing
pions. On the other hand, from the pole diagrams we obtain
\begin{equation}
{\cal B}_{pole}(B \rightarrow B' \pi^0) =
-2 \sqrt{2}\ \overline{M} \ \sum_{B''} \left[
     g_A^{B'B''} \frac{{\cal A}(B \rightarrow B'' \pi^0)}{M_B - M_{B''}} +
     g_A^{B B'}  \frac{{\cal A}(B'' \rightarrow B' \pi^0)}{M_{B'} - M_{B''}}
                                                \right] \ ,
\label{bpole}
\end{equation}
where we have neglected the small $K$ pole term contribution in
Fig.\ 1d, and we have made use of the generalized Goldberger-Treiman
relations to write the strong coupling constants in terms of the axial
charges. Notice that Eq.\ (\ref{bpole}) includes a sum over a set
of intermediate states $B''$. In our calculations, we have included
the $J^\pi=\frac{1}{2}^+$ collective eigenfunctions that arise from
the exact diagonalization of the strong Hamiltonian. It turns out that
only a few excited states are needed to find convergence and their
contribution represents, at most, $15 \%$ of the total values of
${\cal B}_{pole}$.

For the sake of consistency, in order to estimate the size of
the ${\cal B}_{pole}$ amplitudes we will take into account both
the axial charges and the ${\cal A}$ amplitudes obtained within
our model. Therefore, we consider the axial charge operator
$\hat g_A$ arising from the action (\ref{action}), which reads
\begin{eqnarray}
 \frac{\hat g_A}{\sqrt{2}} & = & - \frac{1}{3}
  \int d^3r \left\{ f_\pi^2 \left( F' + 2 c \frac{s}{r} \right) +
            \frac{2}{e^2}  \frac{s}{r} \left( c  \frac{s^2}{r^2} +
            \frac{s}{r} F' + c F'^2 \right) \right\} R_{33}
\nonumber \\
& & + \frac{ f_K^2 - f_\pi^2 }{9} \int d^3r \ (1-c)
\left\{ \left[ \frac{2 s}{r} (1+ 2 c) + F' \right]  ( 1 - R_{88} ) R_{33} +
\left[ \frac{2s}{r} - F' \right] R_{83} R_{38} \right\} \nonumber \\
& & + \frac{N_c }{ 36 \pi^2 \Theta_K}  \int d^3r \ (1-c)
\frac{s}{r} \left( \frac{s}{r} - 2 F' \right) d_{3 k k'} R_{3k}
J_{k'}\ ,
\end{eqnarray}
where $\Theta_K$ is the kaonic moment of inertia. It is worth to
mention that this operator leads to a low value for the
neutron beta decay form factor, $g_A\simeq 0.75$,
compared with the experimental result of about 1.25.

Numerical results for both the contact and pole contributions to the
${\cal B}$ amplitudes are given in Table \ref{tapw}. It can be seen that
the absolute values for the total amplitudes are far too small in
comparison with the experimental results. In the case of the pole
contributions, this could be explained in part by an underestimation
of the axial form factors, as suggested by the low value in the case
of the neutron beta decay. In particular, for ${\cal B}(\Lambda^0_0)$,
the contribution obtained from Eq.\ (\ref{bpole}) using the empirical
values of the ${\cal A}$ amplitudes and axial charges is in very
good agreement with the experimental result (see the value
corresponding to the chiral fit in Table \ref{tapw}). In our calculation,
instead, the suppression arising from the somewhat low predictions for the
$S$-wave amplitudes, together with the underestimation of
the axial form factors, conspire to end up with a reduction factor
of about 1/3. In the case of the remaining $P$-wave amplitudes, it is
well known that the usage of phenomenological
values in (\ref{bpole}) does not allow to get a good fit
of the experimental values. In this sense, the contact
contributions have been suggested as a possible novel ingredient to
solve the discrepancy. Our results show, however, that
the values for ${\cal B}_{contact}$ amount at most 1/10
of the empirical ${\cal B}$ amplitudes. Thus, even if the effect
goes in the right direction, the contact contributions appear to be
too small to represent a potential solution for the $S/P$ problem.

To check the dependence of our results on the Skyrme parameter $\epsilon$
we have considered departures from the central value $\epsilon = 4.1$.
We find that the absolute values of the amplitudes tend to increase
as $\epsilon$ increases. The amplitudes which turn out to be
the most sensitive to the variation of $\epsilon$ are those
corresponding to the $\Lambda \to n \pi^0$ process.
For $\epsilon = 4.5$, which already implies an increase of more that $25 \%$
for the $\Delta N$ splitting, we find, for Set A,
${\cal A}(\Lambda^0_0) \approx - 2 \times 10^{-7}$ which is quite close to
the empirical value ${\cal A}(\Lambda^0_0)_{emp} = - 2.37 \times 10^{-7}$.
However, even in this case, the predicted ${\cal B}(\Lambda^0_0)$ is still
more than a factor 2 below the corresponding empirical value. Thus,
we can conclude that the statements above are quite robust
under variations of the only adjustable parameter in our
calculation.

\section{Contributions of non-octet-like components}

As stated in Sec.\ II, the non-leptonic decay amplitudes are
dominated by the octet-like components of the weak effective
lagrangian, hence only these components have been
considered in the previous two sections. On the other hand,
we have also mentioned that this approximation leads to a
vanishing $\Sigma_+^+$ $S$-wave amplitude. In fact, the
experimental value of ${\cal A}(\Sigma_+^+)$, although significantly
smaller than the other measured $S$-wave amplitudes, is found
to be different from zero.
In this section we will investigate whether within the present model
the standard non-octet-like
contributions to the weak effective lagrangian are able to account
for this difference. It is clear that such contributions will also
modify the results obtained in the previous sections for the other
$S$-wave and $P$-wave decay amplitudes. However, the modifications
are, in the worst case, of the same order of magnitude than the
uncertainties involved in the parameters $g$, $g'$ and $g''$ in
the weak effective lagrangian. Therefore, in what
follows we will concentrate only on the $S$-wave $\Sigma_+^+$ decay
amplitude.

The lowest order 27-plet contribution to the weak effective lagrangian
occurs at $p^2$. It can be written as~\cite{KMW90}
\begin{equation}
{\cal L}^{(27)}_2 = c_3 \ t_{cd}^{ab} \ \Tr\left( Q_a^c \ U^\dagger \partial_\mu U \right) \
\Tr\left( Q_d^b \ U^\dagger \partial^\mu U \right)
\label{lw227}
\end{equation}
where $\left( Q^c_a \right)_{ij} = \delta_{cj} \delta_{ai}$, and
\begin{eqnarray}
t_{21}^{31} &=& t_{12}^{13} = t_{13}^{12} = t_{31}^{21} = \frac{3}{2}\ , \nn \\
t_{12}^{31} &=& t_{21}^{13} = t_{13}^{21} = t_{31}^{12} = 1 \ ,
\end{eqnarray}
with $t_{cd}^{ab} = 0$ otherwise. As in the case of the octet-like piece, we
will take into account also the effect of next-to-leading order couplings.
We consider the ${\cal O}(p^4)$ interaction
\begin{eqnarray}
{\cal L}^{(27)}_4 & = & \quad g_1 \ t_{cd}^{ab} \ \Tr\left( Q_a^c\,
L_\mu\right) \ \Tr\left( Q_d^b \ L_\nu\, L^\mu\, L^\nu \right)
\nonumber \\ & & +\, g_2 \ t_{cd}^{ab} \ \Tr\left( Q_a^c \ L_\mu
\right) \ \Tr\left( Q_d^b \ \{L^\mu,L^2\} \right) \nonumber \\
& & +\, g_3 \ t_{cd}^{ab} \ \Tr\left( Q_a^c \ [L_\mu,L_\nu] \right)
\ \Tr\left( Q_d^b \ [L^\mu,L^\nu] \right)\ .
\label{lw427}
\end{eqnarray}
Once again the most general lagrangian allowed by chiral symmetry
includes many possible terms\cite{KMW90}, and the coupling
constants cannot be fully determined from the available
information on the kaon sector. The structure chosen in
(\ref{lw427}) is, however, sufficient to get a good fit
of $2\pi$ and $3\pi$ $K$ decays. From such a fit one obtains\cite{KMW91}
$g_1\simeq 1.0\times 10^{-10}$, $g_2\simeq -0.4\times 10^{-10}$,
$g_3\simeq 0.1\times 10^{-10}$ together with
$c_3/f_\pi^2 \simeq -0.8 \times 10^{-9}$. This set of values is used
in the numerical calculation below.

The desired $S$-wave $\Sigma^+_+$ amplitude can be now easily
obtained by inserting the explicit form of the chiral field $U$
in the effective couplings (\ref{lw227}) and (\ref{lw427}). By doing
this we arrive to
\begin{equation}
{\cal A}^{(27)}(\Sigma^+ \rightarrow n\,\pi^+) =
\frac{\sqrt{15}}{8\sqrt{2}}\; \left( I_2\, +\,
I_4\right)
\; \langle n | D^{27}_{-{3\over2},0} |\, \Sigma^+ \rangle\ .
\label{a27}
\end{equation}
Here, the left lower index of the SU(3) Wigner function $D$ stands
for $(Y,I,I_3) = (1,3/2,-3/2)$ while the right lower index
corresponds to $(0,0,0)$. The radial integrals $I_2$ and $I_4$ are
given by
\begin{eqnarray}
I_2 & = & -\frac{16 \pi}{3f_\pi}\, c_3 \int dr \ r^2
\left( F'^2 + 2 \ \frac{\sin^2 F}{r^2} \right) \nonumber \\
I_4 & = & -\frac{16\pi}{3f_\pi} \int dr \ r^2 \left[ g_1\,
F'^2\left( F'^2 - 4 \ \frac{\sin^2 F}{r^2} \right) +
2 g_2 \left( F'^2 + 2 \ \frac{\sin^2 F}{r^2} \right)^2  \right. \nonumber
\\
& & \qquad \qquad \qquad \quad \left. +\,
g_3 \,\frac{8\,\sin^2 F}{r^2}
\left( 2\, F'^2 + \frac{\sin^2 F}{r^2} \right) \right]\ .
\end{eqnarray}

Even if the integrand of $I_4$ is suppressed by
the coefficients $g_i$ (which are one to two orders of magnitude lower
that $c_3/f_\pi^2$), it can be seen that the suppression is compensated
by the values of the radial integrals, in such a way that at the
end $I_4$ dominates over $I_2$. Evaluating the matrix element
in (\ref{a27}), we finally obtain
\begin{equation}
{\cal A}(\Sigma^+_+) \simeq 0.01 \times 10^{-7}\ ,
\end{equation}
to be compared with the empirical value
${\cal A}_{emp}(\Sigma^+_+) = 0.13 \times 10^{-7}$ given in
Table \ref{tasw}. We observe
that our estimation for ${\cal A}(\Sigma_+^+)$, although non-vanishing,
is roughly one order of magnitude smaller than the empirical result.
As in the case of the octet-like contributions this statement
remains valid for reasonable variations of the Skyrme parameter
around its central value $\epsilon = 4.1$.

\section{Conclusions}

In this work we have revisited the problem of the calculation of the non-leptonic hyperon decay amplitudes in the
topological soliton models. We have used the approach to the $SU(3)$ Skyrme model in which both the isospin and the
strange degrees of freedom are treated as collective rotations around the usual $SU(2)$ hedgehog ansatz and the
symmetry breaking terms in the strong action are diagonalized exactly. To describe the weak interactions we have
used a chiral effective action, in which low energy constants are adjusted to describe the known $2 \pi$ and $3
\pi$ weak kaon decays. For the $S$-wave decay amplitudes we have found that, compared with previous calculations
based on effective weak chiral lagrangians\cite{DGL85}, the use of empirical input parameters in the strong
effective action, together with the exact diagonalization of the $SU(3)$ collective hamiltonian, lead to a
significant improvement in the predictions. A similar result has been recently obtained using a Cabbibo
current-current type weak interaction\cite{Sco98}. Although our predictions are about $30 \%$ below the empirical
values we consider them as satisfactory in view of the simplicity of the model and the fact that higher order $N_c$
corrections of that size are to be expected. On the other hand, our results badly fail to reproduce the empirical
$P$-wave amplitudes. In soliton models, such amplitudes receive two types of contributions\cite{DGL85}, namely
those arising from the usual pole diagrams and those coming from contact terms. The presence of the latter provided
some hope that the long standing $S/P$ wave puzzle could find a solution within these models. Our results show
that, unfortunately, such contact contributions are far too small to close the gap between the predictions coming
from the pole terms alone and the empirical values. Although one cannot exclude some corrections to these results due to
higher order effects neglected in this work (such as e.g. the kaon induced components which are known to play a
significant role in the determination of the parity violating $\pi N$ coupling constant\cite{MW99}), it is
difficult to believe that they could lead to a solution of this problem. Finally, we have estimated the
contribution to the decay amplitudes coming from non-octet terms in the weak effective action. Since these
contributions are generally very small we have concentrated only on the $S$-wave $\Sigma_+^+$ decay amplitude
which, as well known, vanishes if only octet terms are considered. Our result, although non-zero, turns out to be
roughly one order of magnitude smaller than the empirical value. This clearly indicates that, within the Skyrme
model, more refined wave functions and/or effective weak interactions are needed to understand the subtle effects
related with the small violations of the $\Delta I =1/2$ rule observed in the non-leptonic hyperon $S$-wave decays.

\acknowledgments

D.G.D. acknowledges a Reentry Grant and a research fellowship
from Fundaci\'on Antorchas, Argentina.
This work was supported in part by the grant PICT 03-00000-00133
from ANPCYT, Argentina. N.N.S. is fellow of CONICET, Argentina.

\appendix

\section*{Calculation of the collective matrix elements}

As explained in the main text the calculation of the
decay amplitudes involves the matrix elements of some collective operators
$\hat {\cal O}$ between baryon wavefunctions. To evaluate these matrix
elements we proceed as follows.

In general, the wavefunction corresponding to a baryon $B$ can be expanded
in terms of $SU(3)$ Wigner functions $D^R_{\alpha \beta}$,
\begin{equation}
\Psi_B = \sum_R C_B^R\, \sqrt{\mbox{dim\,}(R)}\,
D_{\alpha\beta}^R \;,
\end{equation}
where $\alpha=(Y,I,I_3)$ and $\beta=(1,J,J_3)$ carry the baryon
quantum numbers, and $R$ is the corresponding representation. The
coefficients $C_B^R$ are obtained from the diagonalization
procedure described in Sec.\ II.
In addition, the collective operators $\hat {\cal O}$ can always be
expressed as
\begin{equation}
\hat {\cal O} = \sum_{\hat\alpha,\hat\beta,\hat R}
\gamma_{\hat\alpha,\hat\beta}^{\hat R}\;\;
D_{\hat\alpha,\hat\beta}^{\hat R}\;,
\end{equation}
where $\gamma_{\hat\alpha,\hat\beta}^{\hat R}$ are numerical
coefficients. These coefficients result from expressing
the ``cartesian" $SU(3)$ indexes $a=1...8$ in terms
of the ``spherical" $SU(3)$ indexes $(Y,I,I_3)$ and performing
suitable Clebsch-Gordan series expansions when needed.
For example, for the operator $R_{78}$ appearing
in Eq.\ (\ref{amp}) we have
\begin{equation}
R_{78} = \frac{i}{\sqrt{2}} \left[D^8_{(1,\frac{1}{2},-\frac{1}{2}),(0,0,0)} -
D^8_{(-1,\frac{1}{2},\frac{1}{2}),(0,0,0)}\right]\;,
\end{equation}
while the combination $R_3^{(+)}$ in Eq.\ (\ref{rpm}) can be written as
\begin{eqnarray}
R^{(+)}_{3} & = &\frac{\sqrt{6}}{10}
\left[
D^8_{(1,\frac{1}{2},-\frac{1}{2}),(0,1,0)} +
D^8_{(-1,\frac{1}{2},\frac{1}{2}),(0,1,0)}
\right] +
\frac{1}{10}
\left[
D^{27}_{(1,\frac{1}{2},-\frac{1}{2}),(0,1,0)} +
D^{27}_{(-1,\frac{1}{2},\frac{1}{2}),(0,1,0)}
\right] \nonumber \\
& & \qquad \qquad - \frac{1}{\sqrt{5}}
\left[
D^{27}_{(1,\frac{3}{2},-\frac{1}{2}),(0,1,0)} +
D^{27}_{(-1,\frac{3}{2},\frac{1}{2}),(0,1,0)}
\right]
\end{eqnarray}

Thus, the matrix element $\langle B' |\hat {\cal O}| B \rangle$ can finally be
expressed in terms of the standard $SU(3)$ Clebsch-Gordan coefficients\cite{Swa63}.
Namely,
\begin{equation}
\langle B' |\hat {\cal O}|B \rangle =
\sum_{\hat\alpha,\hat\beta,\hat R}
\gamma_{\hat\alpha,\hat\beta}^{\hat R}\;
\sum_{R,R',\mu} C_B^R\; (C_{B'}^{R'})^\ast\,
\sqrt{\frac{\mbox{dim\,}(R)}{\mbox{dim\,}(R')}}
\left(\begin{array}{ccc}
\hat R & R & R' \\ \hat\alpha & \alpha & \alpha'
\end{array}\right)
\left(\begin{array}{ccc}
\hat R & R & R'\mu \\ \hat\beta & \beta & \beta'
\end{array}\right)\;,
\end{equation}
where the brackets indicate the Clebsch-Gordan coefficients.
The sum over $\mu$ refers to the situations in which the
Clebsch-Gordan expansion of the product of two $D$'s includes
more than one representation with the same dimension.

\pagebreak

\begin{table}
\narrowtext \caption{Calculated $S$-wave non-leptonic hyperon
decay amplitudes as compared with the empirical values. For
comparison we list also the results of the quark model (QM)
calculations of Ref.\ [1]. All values should be multiplied by
$10^{-7}$. The numerical values of the coefficients $g, g'$ and
$g''$ (which appear in the weak effective lagrangian)
corresponding to Set A and B are given in the text.}

\label{tasw}

\begin{tabular}[h]{ccccc}
                &\multicolumn{2}{c}{This calculation} &  QM   &  Emp  \\
                \cline{2-3}
                &  SET A  &  SET B &     & \\ \hline
$\Lambda^0_0$   & -1.63   & -1.28  & -1.5   & -2.37 \\
$\Sigma^+_0$    & -2.48   & -1.94  & -3.8   & -3.27 \\
$\Sigma^+_+$    &    0    &    0   &   0     &  0.13 \\
$\Xi^0_0$       &  2.37   &  1.86  &  3.0   &  3.43 \\
\end{tabular}

\end{table}

\begin{table}
\narrowtext \caption{Calculated $P$-wave non-leptonic hyperon
decay amplitudes as compared with the empirical values. For
comparison we list also the results of a typical chiral fit
($\chi\,$-fit) taken from Ref.\ [18]. All values should be
multiplied by
$10^{-7}$. The numerical values of the coefficients $g, g'$ and
$g''$ (which appear in the weak effective lagrangian)
corresponding to Set A and B are given in the text.}

\label{tapw}
\begin{tabular}[h]{ccccccccc}
                &\multicolumn{6}{c}{This calculation} &
                \hspace{.2cm}$\chi\,$-fit \hspace{.2cm}&  Emp
                 \\ \cline{2-7}
                &\multicolumn{3}{c}{SET A} & \multicolumn{3}{c}{SET B} & &
                \\ \cline{2-4} \cline{5-7}
                & pole    & contact & total  &  pole &  contact  & total  &  &  \\ \hline
$\Lambda^0_0$   & -5.15   &  -0.38  & -5.53  & -4.04 &  -1.46    & -5.50  & -16.0 &  -15.8 \\
$\Sigma^+_0$    &  2.73   &   2.23  &  4.96  &  2.14 &   2.27    &  4.41  &  10.0 &   26.6 \\
$\Sigma^+_+$    & -0.65   &   4.52  &  3.87  & -0.51 &   3.57    &  3.06  &   4.3 &   42.2 \\
$\Xi^0_0$       &  1.98   &  -0.27  &  1.71  &  1.55 &  -0.003   &  1.55  &   3.3 &  -12.3 \\
\end{tabular}
\end{table}

\begin{figure}[p]
\centerline{\psfig{figure=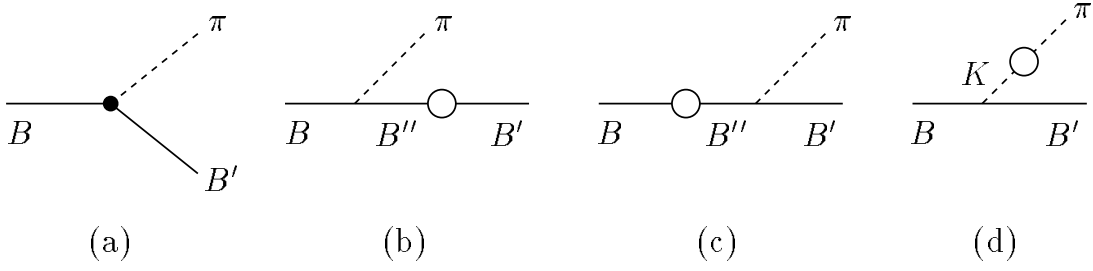,height=6.cm}} \caption[]{Diagrams contributing to the $P$-wave decays: (a)
contact contribution, (b), (c) baryon pole contributions, (d) kaon pole contribution.} \label{f1}
\end{figure}

\end{document}